\documentclass{llncs}
\usepackage{mathptmx}

\usepackage[english]{babel}
\usepackage[pdftex]{graphicx}
\usepackage[T1]{fontenc}

\usepackage{color}
\usepackage{array}
\usepackage[caption=false]{subfig}
\usepackage{url}
\usepackage{cite}

\usepackage{listings}

\newcommand{\todobi}{\begin{itemize} \itemsep0em \footnotesize}
\newcommand{\todoe}{\end{itemize}}

\newcommand{\bm}[1]{\textbf{#1}}

\newcommand{\bi}{\begin{itemize}}
\newcommand{\ei}{\end{itemize}}
\newcommand{\bq}{\begin{equation}}
\newcommand{\eq}{\end{equation}}

\newcommand{\img}[1]{Images/#1}

\newcommand{\mirecv}{\texttt{MPI\_Irecv}}
\newcommand{\mrecv}{\texttt{MPI\_Recv}}
\newcommand{\misend}{\texttt{MPI\_Isend}}
\newcommand{\msend}{\texttt{MPI\_Send}}
\newcommand{\mwait}{\texttt{MPI\_Wait}}

\newcommand{\yes}{Yes}
\newcommand{\no}{No}
\newcommand{\pa}{Partial}
\newcommand{\apsm}{APSM}

\usepackage[
  a4paper,
  colorlinks=true,
  linkcolor=blue,
  citecolor=blue,
  urlcolor=blue,
  bookmarksopen=true,
  bookmarksnumbered=true
]{hyperref}

\title{Asynchronous MPI for the Masses}
\titlerunning{Asynchronous MPI for the Masses}  
\author{
    Markus Wittmann\inst{1} \and
    Georg Hager\inst{1} \and
    Thomas Zeiser\inst{1} \and
    Gerhard Wellein\inst{2}
}
\authorrunning{M. Wittmann et al.} 
\institute{
  Erlangen Regional Computing Center, University of Erlangen-Nuremberg, \\
  Martensstra{\ss}e. 1, 91058 Erlangen, Germany, \\
  \email{hpc@rrze.fau.de},
\and
  High-Performance Computing, University of Erlangen-Nuremberg, \\
  Martensstra{\ss}e 1, 91058 Erlangen, Germany
}

\begin{document}

\mainmatter              

\maketitle               

\begin{abstract}
We present a simple library which equips MPI implementations
with truly asynchronous non-blocking point-to-point operations,
and which is independent of the underlying communication
infrastructure.
It utilizes the MPI profiling interface (PMPI) and the
\texttt{MPI\_THREAD\_MULTIPLE} thread compatibility level, and works
with current versions of Intel MPI, Open MPI, MPICH2, MVAPICH2, Cray
MPI, and IBM MPI. We show performance comparisons on a commodity
InfiniBand cluster and two tier-1 systems in Germany, using low-level
and application benchmarks. Issues of thread/process placement and the
peculiarities of different MPI implementations are discussed in
detail. We also identify the MPI libraries that already support asynchronous
operations. Finally we show how our ideas
can be extended to MPI-IO.
\end{abstract}


\section{Introduction}

\sloppy

A widespread misconception about MPI's non-blocking point-to-point
and I/O routines is that communication and I/O necessarily overlaps with 
computation.
According to the MPI standard \cite{mpi-1.0} non-blocking semantics
does not require asynchronous progress.
Some MPI implementations do support asynchronous progress. However,
this feature often needs to be explicitly enabled at compile time 
(e.g., with ``progress threads''), 
requires the cooperation of several components of the used software stack, or
needs special start-up parameters.

Surprisingly, many applications show performance improvements when non-blocking
point-to-point communication is employed, even when the MPI library does
not feature asynchronous progress. This is because of the other beneficial
consequences of using non-blocking MPI, such as full-duplex transfers
and avoidance of frequent explicit mutual synchronization.

In this work we describe a library which achieves asynchronous 
data transfer by utilizing the profiling interface of MPI (PMPI) and a separate
progress thread.
Therefore the MPI implementation must support the MPI-2 standard \cite{mpi-2.0}
and must provide the \texttt{MPI\_THREAD\_MULTIPLE} compatibility level to 
allow several threads of a process to perform concurrent calls to MPI 
routines.

The library supports the C and Fortran MPI interfaces and requires no code changes
to the target application.
If the target application is statically linked
against the MPI library, relinking may be required.
The method should work independently of the 
underlying interconnect.

This paper is organized as follows: In Sect.~\ref{sec:related-work} we review
related work on asynchronous data transfer with non-blocking point-to-point
MPI or MPI I/O.
The implementation details of the library are discussed in
Sect.~\ref{sec:implementation}, and 
the test bed (hardware and software used for benchmarking) 
is introduced in Sect.~\ref{sec:test-bed}.
In Sect.~\ref{sec:p2p} we evaluate the point-to-point messaging capabilities
of our library using low-level benchmarks and hybrid-parallel
sparse matrix vector multiplication.
Asynchronous I/O is discussed in Sect.~\ref{sec:io}.
Sect.~\ref{sec:summary} gives a conclusion and an outlook.


\section{Related Work}
\label{sec:related-work}

Overlapping data transfer can be achieved on three different
levels \cite{hoefler-2008}:
either by \textit{manual progression}, \textit{progress threads}, or
\textit{communication offload}.\\
\textit{Offloading}~~~
At the lowest level the transfer can be \textit{off\/loaded} to the corresponding
network interface, if supported.
In principle this can be done for example with Myrinet or InfiniBand if the
host channel adapter (HCA) is capable of it.
In \cite{koop-2009} \textsc{Koop} et al. describe a protocol which enables full
asynchronous progress by completely offloading the message transfer and matching
to the InfiniBand HCA.\\
\textit{Progress Threads}~~~
Another option to handle communication and/or I/O while the user application
can proceed is to use dedicated threads.
One option is to have these threads be controlled by the MPI library.
This technique was used for Open MPI until version 1.5.3 \cite{openmpi-web-new};
however, Open MPI could still support threads in some layers of its architecture.
Other implementations, such as MPICH2, MVAPICH2, and Cray MPI \cite{cray-man-mpich2-2012}
also feature
special settings for enabling internal progress threads.
%
%
Also several not so familiar MPI implementations have been built with this idea
in mind, e.g. FiTMPI in \textsc{Mao} et al. \cite{mao-2006} or USFMPI in
\textsc{Caglar} et al. \cite{caglar-2003}.
In \cite{hoefler-2008} \textsc{Hoefler} et al. analyze the impact of progress threads
for non-blocking collectives inside their own reduced MPI implementation.
Their findings are that polling for progress is beneficial if spare cores are available.
Interrupt-driven progress threads are advantageous if all cores are fully utilized.
\textsc{Dickens} et al. \cite{dickens-1999} use progress threads for collective
MPI-IO.
They have found that na\"{\i}ve usage of threads decreases performance.
This might no more be true on current systems as the amount of available cores
has increased.
A similar approach has been followed by \textsc{Patrick} et al. \cite{patrick-2008}
by spawning a thread when the non-blocking I/O functions are called.
The thread then calls the blocking counterpart.
%
%
\textsc{Shahzad} et al. \cite{shahzad-2012} use explicit progress threads inside
an application for performing checkpoints of the application.
Here the performance of the application was only marginally reduced compared to
the case without any checkpointing.
Using application-level progress threads \textsc{Schubert} et al.
\cite{schubert-2011} could significantly improve the performance of
hybrid-parallel sparse matrix-vector multiplication.\\
\textit{Manual Progression}~~~
The idea of manual progression is to repeatedly call
\texttt{MPI\_Test} to check for completion.
Hereby the assumption is made that every call into the library drives the
progress.
An evaluation of how frequent these calls should be performed has been done
by \textsc{Hoefler} et al. in \cite{hoefler-2008-thread}.\\
\textit{Benchmarks}~~~
\textsc{White} and \textsc{Bova} \cite{white-1999} have performed an early
investigation of asynchronous progress in MPI libraries.
In \cite{lawry-2002} \textsc{Lawry} et al. describe a benchmark for detecting possible
overlap.
The Sandia MPI Micro-Benchmark Suite \cite{smb} contains a component that measures the
host processor overhead during non-blocking MPI send and receive operations.
The overhead introduced with using the \texttt{MPI\_THREAD\_MULTIPLE} level
was analyzed by \textsc{Thakur} et al. in \cite{thakur-2009}.
Depending on the implementation quality of the library the overhead ranges from
negligible to large.

\section{Solution}
\label{sec:implementation}

The \apsm\ library (Asynchronous Progress Support for MPI) 
is designed to work with every MPI library and any 
interconnect, as long as following conditions are met:
\bi
  \item Every call into the MPI library drives the internal progress engine.
  \item The MPI library supports \texttt{MPI\_THREAD\_MULTIPLE}.
  \item The MPI library supports the MPI profiling interface (PMPI), i.e., for
    every relevant MPI symbol (\texttt{MPI\_Xxx\ldots}) there is a corresponding symbol
    \texttt{PMPI\_Xxx\ldots}. A library can then implement the MPI functions it 
    wants to intercept and can then call the corresponding \texttt{PMPI} routine. 
\ei
The PMPI interface is used to intercept all MPI calls that are 
relevant for non-blocking point-to-point messages or non-blocking MPI-IO. 
\begin{figure}[tbp]
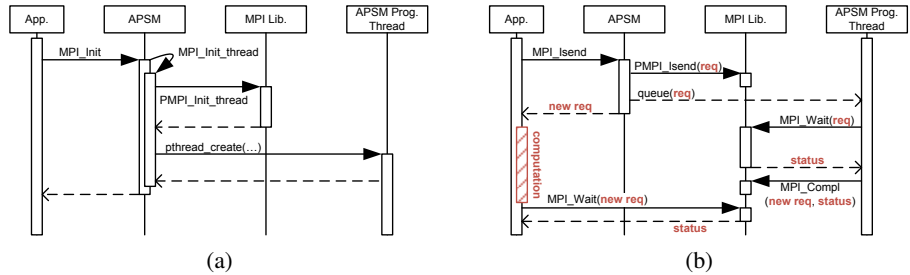

    \subfloat[]{
      \label{fig:async:init}
      \includegraphics[width=0.46\textwidth, clip=true]{\img{Flow-Mpi_Init}}
    } \hfill
    \subfloat[]{
      \label{fig:async:isend}
      \includegraphics[width=0.46\textwidth, clip=true]{\img{Flow-Mpi_Isend}}
    } \,
    \caption{
      \protect\subref{fig:async:init} 
      The \apsm\ intercepts the initialization stage of the MPI 
      library and enforces
      the \texttt{MPI\_THREAD\_MULTIPLE} level.
      Finally the progress thread is started.
      \protect\subref{fig:async:isend} The non-blocking MPI calls 
      (e.g., \texttt{MPI\_Isend}) are intercepted and executed by the 
      application thread.
      The returned request handle is put into an internal queue, which is consumed
      by the progress thread.
      A \textit{generalized request handle} is returned to the application.
      This handle will receive the status of the original request once it has been 
      completed.
    }
    \label{fig:async}
\end{figure}

\subsection{Initialization and Finalization}

The initialization process is depicted in Fig.~\ref{fig:async:init}.
To setup the library the \texttt{MPI\_Init*} functions are intercepted.
In \texttt{MPI\_Init\_thread} the \texttt{MPI\_THREAD\_MULTIPLE} level
is requested; if the library does not support this level, the application is aborted.
(For convenience calls to \texttt{MPI\_Init} are rerouted to
\texttt{MPI\_Init\_thread} so that applications which do not call the
threaded initialization require no code changes.)
Next, the progress thread is created using \verb.pthread_create..
In our experience the Pthreads primitives used by \apsm\ do not interfere 
with any other threading model employed in the user program, such as OpenMP.

The progress thread is terminated by intercepting \texttt{MPI\_Finalize},
which first stops the progress thread before calling \texttt{PMPI\_Finalize}.

\subsection{MPI Point-to-Point Functions}
All intercepted non-blocking point-to-point functions are handled in the
same way (see Fig.~\ref{fig:async:isend}).
If such a function is called by the application the actual
requested operation, e.g., \texttt{MPI\_Isend}, is carried out by calling
the corresponding PMPI function.

The returned request handle is enqueued to an internal queue, and
a newly crated \textit{generalized request handle} is returned to the
application. This handle will act as a ``proxy'' of the original request.
%
%
This process is transparent to the application, and no code changes are
necessary.

The queued original requests which are bound to the message transfers 
are processed from the internal queue by the progress thread.
If, at any time, a bunch of requests is waiting in the queue,
they are served simultaneously by calling either
\texttt{MPI\_Test(some|any)} or \texttt{MPI\_Wait(some|any)}
for driving the progress of data transfer and
waiting for the completions. Which of the four alternatives is
chosen in practice depends on the MPI library (see below).
If a request completes, its status is propagated to the associated
generalized request, notifying the application.

Since all non-blocking MPI functions return an MPI 
request handle, this method will work for them. 
Note that the call to the PMPI functions happens still in the 
context of the application's thread. 
This is necessary to provide a correct MPI program, since
a pair of matching non-blocking send and receive calls
in the same MPI process are guaranteed to complete.
This would not be ensured if the PMPI calls were done inside the
progress thread and \texttt{MPI\_Wait(some|any)} were used to wait
for completion:
After the application posts the send, the progress thread would detect it in
the internal queue, execute it, and wait for it.
Then, a matching receive could be posted next by the 
application, but it would never be handled by the progress thread,
which would wait forever for the completion of the send request.

\subsection{MPI-IO Functions}

Calls to non-blocking MPI-IO functions are handled slightly differently.
The call to the PMPI function (e.g. \texttt{PMPI\_File\_iwrite}) is 
performed in the context of the progress thread. 
Since the MPI standard allows MPI-IO progress to occur within
the initial non-blocking call, this is the more general (and, in this
case, safe) way to ensure asynchronous I/O.

\subsection{Fortran Interface}

The Fortran interface poses a slight problem, since
different MPI implementations use different strategies
to implement the MPI interface.
Internal MPI routines may be called directly from the Fortran 
interface, or the Fortran interface may be stacked on top of the C interface, 
so that Fortran MPI calls are just rerouted.
The library is built to cope with both situations.
Moreover it detects the symbol convention 
for the Fortran routines automatically: As there is no standard for 
how Fortran compilers name their symbols, the 
C \texttt{MPI\_Isend} function could be called, e.g.,
\texttt{mpi\_isend}, \texttt{mpi\_isend\_}, \texttt{mpi\_isend\_\_}, or 
\texttt{MPI\_ISEND}, to name only the most common ones.


\subsection{Affinity of the Progress Thread}

Since the application and the MPI library are not aware of the progress thread,
they cannot control its affinity (i.e., which logical core it is bound to) in a 
meaningful manner. If there is no general way to handle excess threads (e.g., 
with the \verb.-d. option in Cray's \verb.mpirun.), this can be set by the environment variable
\texttt{MPI\_ASYNC\_CPU\_LIST}.
The specified core list relates to the MPI processes on a node.
For example, the list \texttt{0\_2\_4} would pin the progress thread of the first MPI process on every
node to core 0, the progress thread of the second MPI process to core 2, 
and the progress thread of the third MPI process to core 4.\footnote{There is
the residual problem of
how to determine the number of processes per node. This can be solved in a general
way, but an in-depth description would be out of scope for this work.}


\section{Test Bed}
\label{sec:test-bed}

We have used three cluster systems for our tests:
``Lima'' at the Erlangen 
Regional Computing Center (RRZE) in Erlangen, Germany,
``SuperMUC'' at the Leibniz Supercomputing Center (LRZ) in
Garching, Germany, and ``Hermit'' at the High Performance Computing
Center (HLRS) in Stuttgart, Germany.
Their system parameters can be found in Table~\ref{tab:test-bed}.


\begin{table}[tbp]
  \small
  \centering
    \begin{tabular}{ll|rr!{~~}rr!{~~}rr}
      & & \textbf{Lima} & & \textbf{SuperMUC} & & \textbf{Hermit}  \\
      \hline
      CPU type        &&  Intel Xeon    &  & Intel Xeon   &  & AMD Opteron  \\
                      &&  X5650         &  & EP E5-2680   &  & 6276 \\
      Freq. [GHz]     &&  2.67          &  & 2.70         &  & 2.70 \\
      Cores/Socket    &&  6             &  & 8            &  & 2 $\times$ 8 \\
	  \, SMT threads    &&  2       &       & 2       &      & -- \\
      \hline                      
      Sockets         &&  2             &  & 2            &  & 2 \\
      NUMA domains    &&  2             &  & 2            &  & 4 \\
      \hline                      
      Interconnect    &&  QDR IB        &  & FDR10 IB     &  & Cray Gemini \\
      \, Bandw. [GB/s]    &&  $3.2$     &      & $4.2$    &      & $4.6$ \\
      \hline                      
      Parallel FS      &&  Lustre        &  & IBM GPFS     &  & Lustre \\
      \, BW [GB/s]    &&  $\approx$ 3   &  & $\approx$ 200&  & $\approx$ 150
    \end{tabular}
  \caption{Test bed for evaluation.
	For measuring the attainalbe interconnect bandwdith
	the ping-pong part of the Intel MPI benchmarks was used.}
  \label{tab:test-bed}
\end{table}


\section{Overlap of Point-to-Point Messages}
\label{sec:p2p}

In this section we demonstrate the capabilities of the \apsm\ library
using simple low-level benchmarks and a hybrid-parallel sparse
matrix-vector multiplication kernel. Since it is impossible to show
all benchmark results due to the vast number of parameters, we
will concentrate the discussion on the most prominent aspects.
The complete set of benchmark results can  be reviewed online~\cite{apsmresults}.

\subsection{Simple Overlap Benchmark}\label{sec:sol}

A simple benchmark is used to test the ability of MPI libraries
to overlap computation with communication using non-blocking
point-to-point calls \cite{hager-2010}.
Two MPI processes are run, each on its own compute node.
The first process initiates a non-blocking send (\misend{}), performs (CPU-bound)
work for a time $t_w$, and finally calls \mwait{}.
The total time $t_t$ taken for all three steps is measured.
The second process immediately posts a blocking receive (\mrecv{}).
This can also be varied with \mirecv{}/\msend{} or \misend{}/\mirecv{} instead
of \misend{}/\mrecv{}.

When $t_t$ is plotted against $t_w$ and no asynchronous progress has occurred
a straight line can be seen with an offset on the $y$ axis:
\bq
t_t = t_c+t_w ~,
\eq
where the communication time $t_c = V/{B_N}+t_l$. Here $V$ is the message volume, 
$B_N$ is the network bandwidth, and $t_l$
is the network latency.
If there is fully  asynchronous progress, we have
\bq
t_t = \max\left(t_c,t_w\right)~. 
\eq

\begin{figure}[tbp]
    \subfloat[Overlap benchmark w/ Intel MPI 4.0.3]{
      \label{fig:results:ol:impi}
      \includegraphics[width=0.44\textwidth, clip=true]{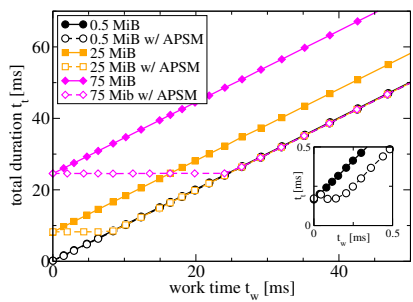}
    } \hfill
    \subfloat[Ping-Pong benchmark w/ Open MPI 1.6.3]{
      \label{fig:results:pp:ompi}
      \includegraphics[width=0.46\textwidth, clip=true]{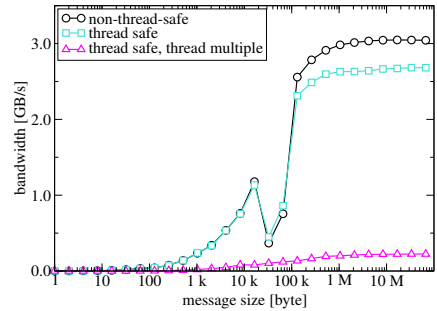}
    }
  \caption{\protect\subref{fig:results:ol:impi} Overlap benchmark  with Intel MPI on
    Lima for \texttt{MPI\_Isend}/\texttt{MPI\_Recv} pair over IB
    with (dashed line) and without (straight line) \apsm. The chosen message size
    is a parameter.
    \protect\subref{fig:results:pp:ompi} PingPong benchmark with Open MPI on Lima.
    Compiling Open MPI with thread safety support introduces a small overhead
    (circles vs. squares).
    When the thread level \texttt{MPI\_THREAD\_MULTIPLE} is additionally requested,
    the InfiniBand transfer module fails to load as it is not thread safe, and
    Open MPI falls back to TCP/IP (triangles).}
\end{figure}

The results for Intel MPI 4.0.3 are shown in Fig.~\ref{fig:results:ol:impi}. 
Intel MPI currently does not provide asynchronous progress over InfiniBand.
Using it together with \apsm\ provides full
overlap for non-blocking point-to-point communication and ``large'' messages
(see below for details on how smaller messages must be handled).
The only test/wait function of MPI which was usable in the progress thread without
deadlocking was \texttt{MPI\_Waitany}. We attribute this to problems with thread safety
in Intel MPI.


Open MPI (version 1.6.3) provides overlap for non-blocking
point-to-point communication, at least for \texttt{MPI\_Isend}.
However, Open MPI cannot use InfiniBand with the \texttt{MPI\_THREAD\_MULTIPLE}
threading level as the corresponding OpenIB module is not thread safe.
In this case the implementation falls back to TCP, which in our case
takes place using IP-over-IB or Gigabit Ethernet.
This can be seen from the simple ping-pong benchmark in Fig.~\ref{fig:results:pp:ompi}.
Here only around $200$ MB/s compared to the $3.0$ GB/s with IB can be achieved
when the highest threading level is requested.

MPICH2 only supports Gigabit Ethernet (GE) and no InfiniBand.
With the internal progress thread enabled (by setting\texttt{MPICH\_ASYNC\_PROGRESS=1})
overlap can be achieved. \apsm\ can be used as an alternative.

MVAPICH2 (version 1.9a2) can overlap non-blocking point-to-point messages with
communication if the internal progress threads are enabled via
\texttt{MPICH\_ASYNC\_PROGRESS=1}.
This MPI library also works with \apsm.

IBM MPI (version 1.2) can by default only overlap \texttt{MPI\_Isend} with computation.
However specifying \texttt{MP\_CSS\_INTERRUPT=yes} \cite{ibm-per-manual-2012},
which causes arriving packets to generate interrupts, leads to overlapping behavior
in all other situations.
Utilizing \apsm\ delivers in principle the same result, but introduces
a lot of variability in execution times; sporadically, MPI calls take
an exceedingly long time. The reason for this behavior has not been
investigated yet.

Cray MPI in the standard configuration provides no asynchronous message 
transfer, but it supports an option to activate an extra progress thread
by setting the environment variable
\texttt{MPICH\_NEMESIS\_ASYNC\_PROGRESS=1} \cite{cray-man-mpich2-2012}
and reserving one core with the \texttt{aprun} option \texttt{-r 1} for the
additional thread.
With the simple overlap benchmark this yields better results
than with \apsm.

Table~\ref{tab:OLsummary} summarizes these results in columns 3 and 5.

\begin{table}[tbp]
  \centering
  \begin{minipage}{\textwidth}\centering
  \begin{tabular}{lc!{~~}c!{~~}c!{~~}c!{~~}cc!{~~}c}
	\textbf{Name}	& \textbf{Version} &  \multicolumn{2}{c}{\textbf{Overlap}} & \textbf{~~Works w/~~} & \multicolumn{2}{c}{\textbf{Improvement}} & \textbf{System} \\
	\cline{3-4} \cline{6-7}
			& & Com. & I/O & \bfseries\apsm & Com. & I/O & \\
	\hline
	Intel MPI 		& 4.0.3
					& \no                                         & \no
					& \pa\textsuperscript{a}
					& \yes                                        & \no\textsuperscript{b}
					& Lima \\
	Open MPI 		& 1.6.3
					& \pa\textsuperscript{c}                      & \no
					& \pa\textsuperscript{d} 
					& \pa\textsuperscript{d}                      & \pa\textsuperscript{d}
					& Lima \\
	MPICH2   		& 1.5
					& \yes\textsuperscript{e}                     & -
                    & \yes
                    & \yes & -
                    & Lima \\
	MVAPICH2 		& 1.9a3
					& \yes\textsuperscript{e} & \no
                    & \yes
                    & \yes                                        & \yes
                    & Lima \\
	IBM MPI 		& 1.2
				    & \yes\textsuperscript{f}            & -
                    & \yes
                    & \yes\textsuperscript{g} & -
                    & SuperMUC \\
	Cray MPI 		& 5.6.1
					& \yes\textsuperscript{e}  & \no
                    & \yes
                    & \yes\textsuperscript{g} & \yes
                    & Hermit \\
 \\
  \end{tabular}
  \caption{\label{tab:OLsummary}
    Overview of all MPI implementations and the system they were evaluated on.
  }\end{minipage}
\footnoterule\footnotesize
\begin{minipage}[t]{0.49\linewidth}
\textsuperscript{a} Only \texttt{MPI\_Waitany} can be called inside the progress thread\par
\textsuperscript{b} I/O deadlocks\par
\textsuperscript{c} Only \texttt{MPI\_Isend} overlaps\par
\end{minipage}\hfill
\begin{minipage}[t]{0.49\linewidth}
\textsuperscript{d} No IB with \apsm\par
\textsuperscript{e} With progress thread enabled\par
\textsuperscript{f} With interrupts enabled\par
\textsuperscript{g} Strong variability in measurement\par
\end{minipage}
\end{table}

\subsection{Prototype Ghost Cell Exchange Benchmark}

This benchmark simulates strong scaling of an application which
performs exchange of ghost cells in one dimension. A number of MPI
processes are running, each of which exchanges a ``halo'' of fixed
size with its two neighbors. After the exchange, each process executes
a workload which is subject to strong scaling with the number of
processes. Computations which would be required in a real application
for the boundary cells in preparation of the halo exchange are
neglected.

In order to better mimic the execution behavior of real applications but still
achieve good reproducibility of time measurements,
a simple triad loop benchmark (\texttt{a(:) = b(:) * c(:) + d(:)})
was chosen as the workload. The size of the working set was adjusted to fit 
into each core's own L2 cache.

We used the Lima cluster with Intel MPI and up to twelve nodes, with twelve 
MPI processes per node (PPN).
Each process was bound to its own physical core. 
In the case where the \apsm\ library was used, the progress thread was pinned to the
other SMT thread. 

\begin{figure}[tbp]
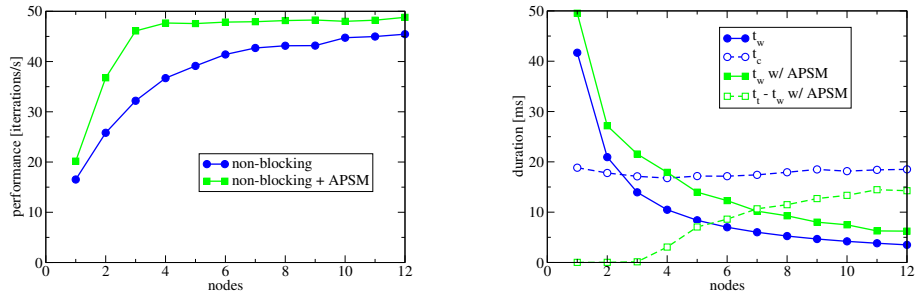

    \subfloat{\label{fig:halo:perf}
      \includegraphics[width=0.44\textwidth, clip=true]{\img{GhostCellBenchmark-Lima-IntleMpi}}
    } \hfill
    \subfloat{\label{fig:halo:bd}
      \includegraphics[width=0.44\textwidth, clip=true]{\img{GhostCellBenchmark-Lima-IntleMpi-breakdown}}
    }
  \caption{
    Ghost cell benchmark on Lima with Intel MPI for one to twelve nodes 
    and twelve MPI processes per node.
    \protect\subref{fig:halo:perf} 
    Performance with plain Intel MPI (circles) and with \apsm\ (squares). 
    \protect\subref{fig:halo:bd} Breakdown of computation ($t_w$) and (visible) 
    communication times.
  }
  \label{fig:gc-bench:intel}
\end{figure}

The performance results for a communication buffer size of $10$~MiB 
can be found in Fig.~\ref{fig:halo:perf}.
Use of the \apsm\ library achieves superior performance and
scalability up to the point where communication takes longer
than computation (at about three nodes). This can be seen from
Fig.~\ref{fig:halo:bd}, which shows a breakdown of time contributions
in the duration of work, i.e. the computation (filled symbols), and the ``visible''
communication time, which is in case of overlap the difference between
overall time and working time (open squares). Without overlap,
the communication time is independent of the number of processes
(open circles), since the message length is always the same.

Beyond three nodes, performance saturates in the overlapped case,
whereas it continues to rise without overlap. At large node counts,
both numbers coincide, since communication absolutely dominates
in this case and computation time is negligible. However,
this is not the limit in which one would want to run any real
application code in practice, since parallel efficiency has
dropped to unacceptable levels. The ``sweet spot'' in terms of
efficient execution is at the point where the overlapped performance
saturates. This is also where the advantage compared to the non-overlapped
case is at a maximum.

As the difference in $t_w$ between overlapped and non-overlapped cases
shows, the progress thread requires
additional resources, reducing the worker thread's performance accordingly. 
In cases where spare physical cores are available, e.g., if the application
is strongly memory-bound with saturation across the cores of a socket,
the progress threads can be bound to those. This would strongly reduce the
interference of MPI progress with application execution.



\subsection{Sparse Matrix Vector Multiplication (spMVM)}

We use a hybrid (OpenMP + MPI) sparse matrix vector
multiplication ($y=y+M \times \bm{v}$) as a relevant real-world test
case to demonstrate the applicability of our approach.
MPI parallelization is done by distributing the
matrix rows across processes so that each process has
(approximately) the same number of nonzero entries. The
right-hand side (RHS) vector $\bm v$ is distributed in the same way.
Consequently, overlapping computation with the required
communication of the RHS vector parts requires splitting
the spMVM operation in two phases: A ``local'' phase, in which
a process multiplies its local part of the RHS vector to
the corresponding diagonal block of the matrix, and a ``non-local''
phase, in which the parts of the RHS vector that have been
received by other processes are multiplied to the remaining
matrix entries.

There are two ways in which communication overlap may be achieved,
``vector model with naive overlap'' and ``task mode with explicit
overlap'' \cite{schubert-2011}. The former uses all OpenMP threads to perform
the local
spMVM part and relies on non-blocking MPI calls and a subsequent
\texttt{MPI\_Waitall} for asynchronous MPI progress, while the latter
uses a dedicated communication thread. After
the local spMVM and the communication are both over, both approaches
perform the non-local spMVM with all threads
(see \cite{schubert-2011} for a full description).

If the MPI implementation does not support asynchronous progress,
communication only takes place during the \texttt{MPI\_Waitall} call.
This results in overlap only with task mode, at the price of
sacrificing one worker thread for the local spMVM.


For evaluation we selected two sparse matrices ``HV15R'' and ``DLR1,'' 
where the former has about $2\cdot 10^6$ rows and $2.8\cdot 10^8$
nonzeros, and the latter has $2.8\cdot 10^5$ rows and $4\cdot 10^7$
nonzeros. DLR1 uses only around $480$~MB of memory and thus fits 
completely in the L3 caches of $24$ Lima compute nodes.
\begin{figure}[tbp]
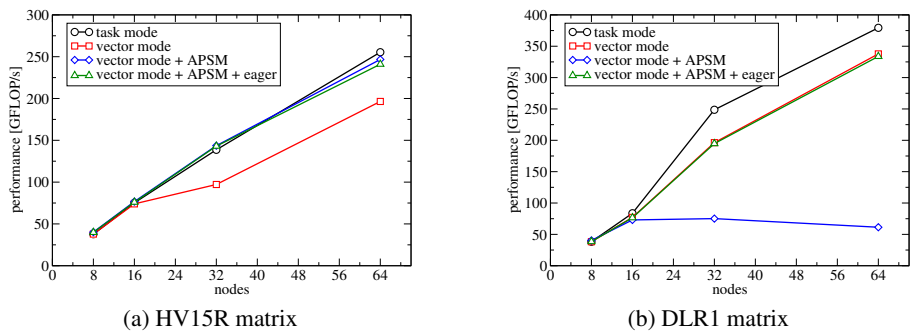

    \subfloat[HV15R matrix]{
      \label{fig:results:spmvm:hv15r:intel}
      \includegraphics[width=0.44\textwidth, clip=true]{\img{spmvm-hv15r-intel}}
    } \hfill
    \subfloat[DLR1 matrix]{
      \label{fig:results:spmvm:dlr1:intel}
      \includegraphics[width=0.44\textwidth, clip=true]{\img{spmvm-dlr1-intel}}
    } \,
  \caption{Performance comparison of spMVM with Intel MPI on Lima, running one
    MPI process with twelve OpenMP threads per node}
  \label{fig:results:spmvm}
\end{figure}

Task mode shows best performance with Intel MPI in all cases (see Fig.~\ref{fig:results:spmvm}).
For the HV15R matrix (Fig.~\ref{fig:results:spmvm:hv15r:intel})
vector mode with \apsm\ is better than without
and nearly achieves the performance of task mode.
The DLR1 matrix reveals a specific problem due to the very small message
sizes that occur as the number of MPI processes is increased.
These are usually handled by MPI implementations using a so-called
\textit{eager} protocol: If a message is small enough
it can directly be sent to a predefined buffer at the destination. 
Otherwise the sender and the receiver must
synchronize to initiate the actual transfer (\textit{rendezvous} protocol).

If  every message were handled in the same way by the \apsm\ library independent of size,
the latency for potential eager messages would increase since the whole
mechanism of processing queued requests by the progress thread would add nothing
but overhead.
When the library is made aware of the threshold, request handles for
eager messages are directly obtained from MPI and passed back to the application,
with no interference from the progress thread.
The difference between both behaviors can be seen in Fig.~\ref{fig:results:spmvm:dlr1:intel},
where the performance of \apsm\ without eager awareness becomes unacceptable
beyond 16 nodes (diamonds).
With eager awareness, however (at a message size of $\leq256$~KiB in this case),
the performance reaches the level of vector mode without \apsm\ (triangles). The reason
why task mode is still measurably better at large node counts is that 
all overheads connected with MPI communication can be hidden by an explicit
communication thread, whereas the eager protocol alone (which is also
in effect when \apsm\ is used with eager awareness) still suffers from
unavoidable communication latencies.


\section{Overlap of MPI-IO}
\label{sec:io}

The \apsm\ library can also be used to overlap computation and MPI-IO.
To evaluate the state of the MPI implementations and the usefulness
of \apsm\ a modified version of the overlap benchmark (see Sect.~\ref{sec:sol})
was used, where 
point-to-point communication was substituted by I/O via \texttt{MPI\_File\_iwrite}.
Only one process per node was used, writing $6$~GiB
of data to a parallel file system. All processes wrote to the same file. Care
was taken to rule out caching effects, i.e., the measured I/O times
included real disk I/O only.
In general, getting reliable timing for I/O is not easy since the parallel
filesystems are usually under load by other users, 
which leads to fluctuating bandwidths. 

\begin{figure}[tbp]
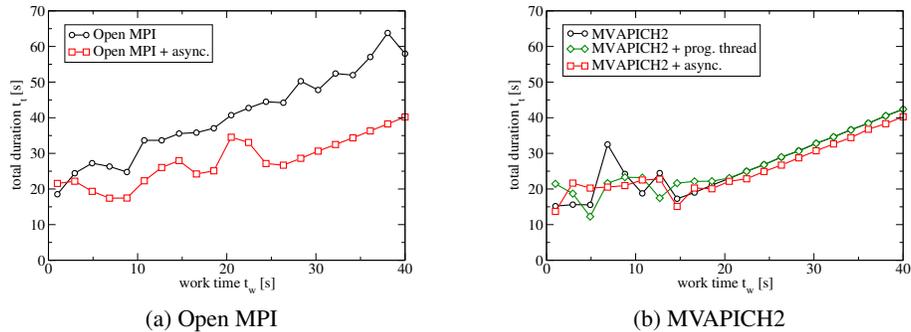

    \subfloat[Open MPI]{
      \label{fig:results:io:openmpi}
      \includegraphics[width=0.44\textwidth, clip=true]{\img{io-openmpi-n-008-6-gib}}
    } \hfill
    \subfloat[MVAPICH2]{
      \label{fig:results:io:mvapich2}
      \includegraphics[width=0.44\textwidth, clip=true]{\img{io-mvapich2-n-008-6-gib}}
    } \,
  \caption{Results for the MPI-IO overlap benchmark on Lima using
    \protect\subref{fig:results:io:openmpi} Open MPI and \protect\subref{fig:results:io:mvapich2} MVAPICH2, 
    on eight Lima nodes, one process per node, and 6 GiB of I/O volume per process.}
  \label{fig:results:io}
\end{figure}

The results can be summarized as follows. 
The Intel MPI library does not overlap computation and I/O.
Using it together with \apsm\ is not possible due to frequent
deadlocks. Open MPI does not provide asynchronous I/O on the Lustre filesystem
of Lima, but overlap can be seen with \apsm\ 
(see Fig.~\ref{fig:results:io:openmpi}).
The main disadvantage with this solution is that Open MPI can not
use native InfiniBand for point-to-point communication with \apsm\ (see Sect.~\ref{sec:sol} 
above).
MVAPICH2 does not support asynchronous I/O even with its internal
progress thread enabled.
However, overlap is available with \apsm\ (see Fig.~\ref{fig:results:io:mvapich2}).
Cray MPI does not feature asynchronous I/O (with and without
the internal progress thread) but can benefit from the \apsm\ library.

Table~\ref{tab:OLsummary} gives an overview of the results in columns 4 and 7.


\section{Summary}
\label{sec:summary}

We have demonstrated how asynchronous point-to-point communication 
and MPI-IO can be achieved for MPI implementations that have no native
support for asynchronous progress. By overloading some MPI functions
using the PMPI interface, an internal progress thread is used to handle
non-blocking requests in the background with minimal impact on the
performance of code execution in the application program. In cases
where no dedicated physical cores are available for the progress thread, 
virtual cores can be used. Most current MPI implementations are
compatible with our method. However, strict \verb.MPI_THREAD_MULTIPLE.
compatibility and thread safety is required. 

Possible future work includes support for persistent communication
and split-collective MPI-IO functions.
The library is freely available under an LGPL license \cite{apsmresults}.

\paragraph{Acknowledgements}
This work was supported by BMBF under grant No.\ 01IH11011C (project FETOL).

\bibliographystyle{abbrv}
\bibliography{Bibliography}

\begin{thebibliography}{10}

\bibitem{caglar-2003}
S.~G. Caglar, G.~D. Benson, Q.~Huang, and C.-W. Chu.
\newblock {USFMPI}: {A} multi-threaded implementation of {MPI} for {L}inux
  clusters.
\newblock In {\em Proceedings of the IASTED Conference on Parallel and
  Distributed Computing and Systems}, 2003.

\bibitem{cray-man-mpich2-2012}
{Cray Corp.}
\newblock Manpage: intro\_mpi(3).
\newblock
  \url{http://docs.cray.com/cgi-bin/craydoc.cgi?mode=Show;q=;f=man/xe_mptm/56/cat3/intro_mpi.3.html},
  2012.

\bibitem{dickens-1999}
P.~M. Dickens and R.~Thakur.
\newblock Improving collective {I/O} performance using threads.
\newblock In {\em Proceedings of the 13th International Parallel Processing
  Symposium and 10th Symposium on Parallel and Distributed Processing}, pages
  38--45, 1999.

\bibitem{hager-2010}
G.~Hager and G.~Wellein.
\newblock {\em Introduction to {H}igh {P}erformance {C}omputing for Scientists
  and Engineers}.
\newblock CRC Press, July 2010.
\newblock ISBN 978-1439811924.

\bibitem{hoefler-2008}
T.~Hoefler and A.~Lumsdaine.
\newblock Message progression in parallel computing - {T}o thread or not to
  thread?
\newblock In {\em Proceedings of the 2008 {IEEE} International Conference on
  Cluster Computing}. {IEEE} {C}omputer {S}ociety, Oct. 2008.
\newblock \bibdoiurl{10.1109/CLUSTR.2008.4663774}.

\bibitem{hoefler-2008-thread}
T.~Hoefler and A.~Lumsdaine.
\newblock Message progression in parallel computing - to thread or not to
  thread?
\newblock In {\em Cluster Computing, 2008 IEEE International Conference on},
  pages 213--222, 2008.
\newblock \bibdoiurl{10.1109/CLUSTR.2008.4663774}.

\bibitem{ibm-per-manual-2012}
{IBM Corp.}
\newblock Improving performance with {MP\_CSS\_INTERRUPT}.
\newblock
  \url{http://publib.boulder.ibm.com/infocenter/zos/v1r13/index.jsp?topic=%2Fcom.ibm.zos.r13.fomp100%2Fipezou0079.htm},
  2012.

\bibitem{koop-2009}
M.~J. Koop, J.~K. Sridhar, and D.~K. Panda.
\newblock Tuple{Q}: {F}ully-asynchronous and zero-copy {MPI} over
  {I}nfini{B}and.
\newblock In {\em Proceedings of the IEEE International Symposium on Parallel
  \& Distributed Processing 2009}, Los Alamitos, CA, USA, 2009. IEEE Computer
  Society.
\newblock \bibdoiurl{10.1109/IPDPS.2009.5161056}.

\bibitem{lawry-2002}
W.~Lawry, C.~Wilson, A.~B. Maccabe, and R.~Brightwell.
\newblock {COMB}: {A} portable benchmark suite for assessing {MPI} overlap.
\newblock In {\em Proceedings of IEEE International Conference on Cluster
  Computing}, pages 472--475, 2002.

\bibitem{mao-2006}
J.~Mao, B.~Song, Y.~Wu, and G.~Yang.
\newblock Overlapping communication and computation in {MPI} by multithreading.
\newblock In {\em Proceedings of The 2006 International Conference on Parallel
  \& Distributed Processing Techniques and Applications}, PDPTA'06, pages
  52--57, 2006.

\bibitem{mpi-1.0}
{Message Passing Interface Forum}.
\newblock {MPI: A Message-Passing Interface Standard, Version 1.0}.
\newblock \url{http://www.mpi-forum.org/docs/mpi-10.ps}, May 1994.

\bibitem{mpi-2.0}
{Message Passing Interface Forum}.
\newblock {MPI-2: Extensions to the Message-Passing Interface}.
\newblock \url{http://www.mpi-forum.org/docs/docs.html}, July 1997.

\bibitem{patrick-2008}
C.~M. Patrick, S.~Son, and M.~Kandemir.
\newblock Comparative evaluation of overlap strategies with study of {I/O}
  overlap in {MPI-IO}.
\newblock {\em SIGOPS Oper. Syst. Rev.}, 42(6):43--49, Oct. 2008.
\newblock \bibdoiurl{10.1145/1453775.1453784}.

\bibitem{smb}
{Sandia National Laboratory}.
\newblock {Sandia MPI Micro-Benchmark Suite (SMB)}.
\newblock \url{http://www.cs.sandia.gov/smb/}.

\bibitem{schubert-2011}
G.~Schubert, H.~Fehske, G.~Hager, and G.~Wellein.
\newblock Hybrid-parallel sparse matrix-vector multiplication with explicit
  communication overlap on current multicore-based systems.
\newblock {\em Parallel Processing Letters}, 21(03):339--358, 2011.
\newblock \bibdoiurl{10.1142/S0129626411000254}.

\bibitem{shahzad-2012}
F.~Shahzad, M.~Wittmann, T.~Zeiser, and G.~Wellein.
\newblock Asynchronous checkpointing by dedicated checkpoint threads.
\newblock In J.~Tr\"aff et~al., editors, {\em Recent Advances in the Message
  Passing Interface}, volume 7490 of {\em LNCS}, pages 289--290. Springer,
  2012.
\newblock \bibdoiurl{10.1007/978-3-642-33518-1_36}.

\bibitem{thakur-2009}
R.~Thakur and W.~Gropp.
\newblock Test suite for evaluating performance of multithreaded {MPI}
  communication.
\newblock {\em Parallel Comput.}, 35(12):608--617, Dec. 2009.
\newblock \bibdoiurl{10.1016/j.parco.2008.12.013}.

\bibitem{openmpi-web-new}
{The Open MPI Project}.
\newblock What's new in {SVN} vs. the current release?
\newblock \url{http://www.open-mpi.org/svn/new.php}.

\bibitem{white-1999}
J.~B. {White III}, , and S.~W. Bova.
\newblock Where's the overlap? -- {A}n analysis of popular {MPI}
  implementations.
\newblock In {\em Third {MPI} Developer's and User's Conference: Proceedings
  {MPIDC}'99, {A}tlanta, {G}eorgia, {M}arch 10--12, 1999}.

\bibitem{apsmresults}
M.~Wittmann.
\newblock Asynchronous progress support for {MPI}.
\newblock \url{https://grid.rrze.uni-erlangen.de/~unrza252/apsm}.

\end{thebibliography}

\end{document}